\documentclass[twocolumn,showpacs,aps,pre,superscriptaddress,floatfix,longbibliography]{revtex4-2}

\usepackage{amsmath,amsfonts,amssymb,amsthm,amstext,mathtools,latexsym}
\usepackage{graphics,graphicx,epsfig,bbm,epsfig,epstopdf,psfrag, bbm}
\usepackage{dcolumn,bm,comment,dsfont,times,dsfont,braket,euscript}
\usepackage{subfigure, setspace}

\usepackage[colorlinks=true,citecolor=blue,linkcolor=blue,urlcolor=blue]{hyperref}
\usepackage[usenames]{color}

\usepackage[utf8]{inputenc}

\UseRawInputEncoding
\begin{document}
%\setstcolor{red}

\title{Emergence of beating in a magnetic flagellum consisting of active bots}
%\title{Beating by propulsion in the magnetic flagellum}
%\title{Beating by propulsion: The Magnetic Flagellum}
%% List all authors/contributors using separate \author macros, and mark the speaker/presenting author (only one) with the *
%% No comma is required

%\date{\today}

\author{Francisca Guzm\'an-Lastra}
\email{fguzman@uchile.cl}
\affiliation{Departamento de F\' isica, Facultad de Ciencias, Universidad de Chile, Santiago Chile}
%\email{denisse.pasten.g@gmail.com}
%\affiliation{Departamento de F\'ısica, Facultad de Ciencias, Universidad de Chile, Santiago Chile}
\author{Daniel Hern\'andez} %\orcid{0000-0001-5376-8062}
\affiliation{Departamento de F\' isica, Facultad de Ciencias, Universidad de Chile, Santiago Chile}

\author{Nicol\'as Quintriqueo}
\affiliation{Departamento de F\' isica, Facultad de Ciencias, Universidad de Chile, Santiago Chile}
%\email{denisse.pasten.g@gmail.com}
%\affiliation{Departamento de F\'ısica, Facultad de Ciencias, Universidad de Chile, Santiago Chile}

\author{Enkeleida Lushi}
%\email{lushi@softactivematterlab.com}
\affiliation{Soft Active Matter Lab, Branchburg, NJ, 08876, United States}

\author{Erick Burgos}
\affiliation{Departamento de F\' isica, Facultad de Ciencias, Universidad de Chile, Santiago Chile}
%\affiliation{Departamento de F´ısica, Facultad de Ciencias, Universidad de Chile, Santiago Chile}

\begin{abstract}
We investigate the emergence of flagellar beating in chains of magnetic 
self-propelled particles (MSPPs) built from centimeter-scale vibrating 
robots (Hexbugs) with embedded neodymium dipoles. When one end of the 
chain is anchored and self-propulsion is activated, longitudinal stress 
accumulates along the chain until it overcomes the magnetic bending 
stiffness, triggering a buckling instability that drives sustained 
flagellar beating. Using a combination of experiments and numerical 
simulations, we identify three distinct dynamical regimes — straight 
chain, stable flagellar beating, and fission — governed by the 
competition between active force, chain length, and magnetic bending 
stiffness. The onset of beating requires a seed misalignment set by 
the balance between magnetic torques and rotational noise, and we show 
that the transition corresponds to a supercritical Hopf bifurcation. 
A kinematic model reproduces the observed orientation dynamics with 
excellent agreement. The magnetic bending stiffness, which arises 
directly from dipole-dipole interactions, is fully tunable via dipole 
strength and chain length, offering independent experimental control 
over both activity and rigidity. Our results establish a macroscopic 
platform for studying force-induced buckling and self-oscillations in 
active filaments, with direct connections to flagellar motion in 
biological and synthetic microswimmers.
\end{abstract}

\maketitle

Among the various interactions governing collective behavior in physical 
systems, magnetic dipole--dipole interactions appear across a broad range 
of scientific and technological contexts, from condensed matter and 
colloidal physics to biophysics and materials 
science~\cite{xie2019reconfigurable,bishop2023active,mccallum2014practical,yang2021motion,meng2019magnetically,khatavkar2007active}. 
In active matter systems, these interactions enable programmable 
self-assembly and long-range orientational 
alignment~\cite{koessel2020emergent,martinez2015colloidal,kiani2015elastic,tierno2008magnetically}, 
with magnetotactic bacteria serving as natural models for directional 
guidance and collective 
behavior~\cite{klumpp2019swimming,thery2024controlling,vincenti2019magnetotactic}, 
while magnetically driven colloids offer synthetic analogs that can be 
precisely controlled via external 
fields~\cite{wang2023spontaneous,xie2019reconfigurable,vutukuri2017rational}. 
These systems have proven effective in applications such as soft robotics, 
microfluidics, and biomedical devices. However, they often exhibit 
dynamic, transient metastable states, making long-term control a 
persistent challenge.

Active magnetic particles have thus emerged as a compelling class of 
systems, both in biological and synthetic contexts, where their complex 
behavior arises from the interplay between self-propulsion, magnetic 
dipolar interactions, and intrinsic 
polarization~\cite{guzman2016fission,martinez2015colloidal,nishiguchi2018flagellar,liao2021emergent,vanesse2023collective}. 
A particularly intriguing case emerges when, in the presence of external 
fields or elastic bonding between particles, the head-to-tail symmetry 
of a linear configuration of active particles is broken, leading to 
flagellar-like beating driven by 
activity~\cite{jayaraman2012autonomous,camalet1999self,sekimoto1995symmetry}. 
This phenomenon, known as beating by propulsion, mirrors the locomotion 
of spermatozoa and eukaryotic 
flagella~\cite{gilpin2020multiscale,werner2014shape,gallagher2019rapid,guasto2020flagellar,ma2014active,cammann2025form,lisicki2024eukaryotic}, 
and has also been observed in synthetic systems such as active filaments and elastohydrodynamic 
simulations~\cite{fily2020buckling,laskar2015brownian,kiani2015elastic,de2017spontaneous,chelakkot2014flagellar,ling2018instability, xi2024emergent, alizzi2026nonreciprocalbuckling}, 
as well as in magnetic Janus particles and non-reciprocal active 
colloids~\cite{gonzalez2018active,vutukuri2017rational,nishiguchi2018flagellar}. 
Recently, this effect has been reproduced at the macroscopic scale using centimeter-scale
Hexbugs: battery-powered robots propelled by internal 
vibrations ~\cite{gutierrez2020inertial,xi2024emergent}. When connected via an elastic membrane, a Hexbug chain spontaneously exhibits oscillatory motion and synchronization, 
demonstrating the universality of this phenomenon across 
scales~\cite{zheng2023self,xia2024biomimetic}. Synchronization between 
flagella and cilia is a fundamental feature of biological locomotion and fluid transport~\cite{wan2024mechanisms}, and has recently been reproduced  in minimal synthetic designs~\cite{moreau2024minimal}.

In this work, we investigate the emergence of beating by propulsion in a chain of inertial, active disk-shaped particles with permanent magnetic dipole--dipole interactions~\cite{musacchio2025fluidization,obreque2024dynamics,sepulveda2021bioinspired}. Particles are arranged in a linear configuration with one end anchored to a rigid wall~\cite{chelakkot2014flagellar}, remaining bound via dipole--dipole interactions. Upon activation, self-propulsion induces longitudinal stress along the chain that, depending on the interplay between activity, magnetic interaction, and orientational 
fluctuations, drives the system toward one of three regimes: a static 
straight chain, stable flagellar-like beating, or 
fission~\cite{guzman2016fission,kaiser2015active}. We develop an 
analytical model for the tangential misalignment that triggers 
buckling~\cite{ling2018instability,jayaraman2012autonomous}, and show 
that beating onset arises from the competition between active force 
$F_0$, dipolar attraction $F_{\rm mag}$, and orientational fluctuations 
$D_R$. We further characterize beating amplitude and frequency as 
functions of magnetic strength and chain length through bending rigidity, bond angles, and three-body bending energy~\cite{kiani2015elastic}, providing a controllable 
platform for exploring active solids and self-oscillating filament 
dynamics~\cite{zheng2023self,tejedor2024progressive}.

\begin{figure}
    \centering 
\includegraphics[width=1.0\columnwidth]{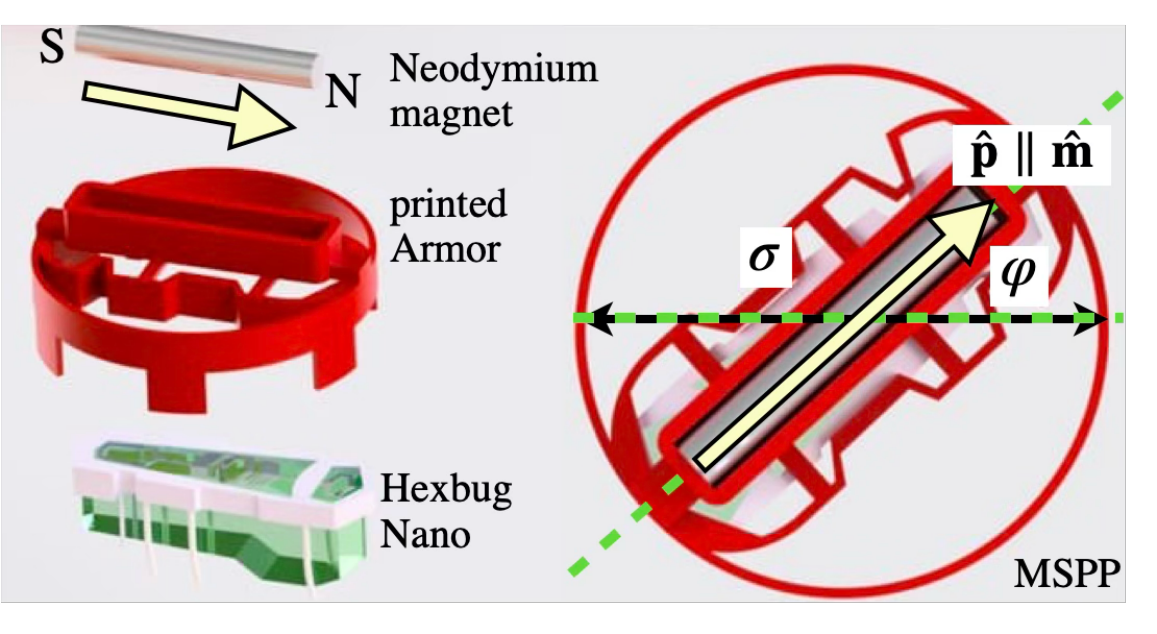}	
    \caption{\textbf{Magnetic self-propelled particle (MSPP).} Left: 
components of an MSPP — a neodymium cylindrical magnet (S--N poles 
indicated), a 3D-printed disk-shaped armor of diameter $\sigma = 5$~cm, 
and a Hexbug Nano robot providing self-propulsion via internal vibratory 
motion. Right: schematic of the assembled MSPP, modeled as a disk of 
diameter $\sigma$ with orientation angle $\varphi$ measured from the 
chain axis (green dashed line, corresponding to the $\hat{\mathbf{x}}$ 
axis). The self-propulsion direction $\hat{\mathbf{p}} = (\cos\varphi, 
\sin\varphi, 0)$ is parallel to the permanent magnetic dipole moment 
$\mathbf{m} = m\hat{\mathbf{p}}$, with magnitude $m = 0.6 \pm 0.1$~Am$^2$ 
determined via gaussmeter measurements~\cite{obreque2024dynamics}.} 
\label{fig1}
\end{figure}

\section{Results and discussion}

\subsection{Experimental details}

We experimentally study the dynamics of a chain composed of $N$ magnetic self-propelled particles (MSPPs), constructed from Hexbug Nano 
robots~\cite{Hexbugs} enclosed within 3D-printed disk-shaped armors of 
diameter $\sigma = 5.0 \pm 0.1$~cm and mass $M = 0.018$~kg. Each disk 
includes a cylindrical compartment along its diameter that houses a 
neodymium rod magnet of diameter $d = 0.6 \pm 0.001$~cm and length 
$L = 3.5 \pm 0.1$~cm, whose permanent magnetic dipole moment 
$m = 0.6 \pm 0.1$~Am$^2$ was determined via gaussmeter 
measurements~\cite{obreque2024dynamics}. Each MSPP self-propels along 
its body axis, oriented at angle $\varphi_i$ with respect to the horizontal, 
with a magnetic dipole moment $\mathbf{m}_i = m\hat{\mathbf{p}}_i$ 
parallel to the propulsion direction (see Fig.~\ref{fig1}).

To ensure uniform propulsion, all experiments were carried out using new batteries, which allowed each MSPP to self-propel at a consistent average speed $u_0^{\text{hexbugs}} = 2$~cm/s~\cite{obreque2024dynamics}, leading 
to an active force $F_0 = \gamma_T u_0^{\text{hexbugs}} = 2\times10^{-4}$~N. 
The translational friction coefficient $\gamma_T = 10^{-2}$~kg/s, rotational friction coefficient $\gamma_R=10^{-3}$~kg m$^2$/s, and rotational diffusion coefficient 
$D_R = 10^{-5}$~rad$^2$/s were not measured directly but were selected to be consistent with the known dynamical behavior of Hexbug-based MSPPs in previous experiments~\cite{obreque2024dynamics,musacchio2025fluidization,zheng2023self}, 
and validated by matching simulation trajectories to experimental 
observations (see Supplemental Material~\cite{SM}). Due to the limited 
self-propulsion speed of the Hexbugs, we focus on chain lengths of $N=4$ 
and $N=5$ MSPPs. Throughout the main text we present results for the $N=4$ 
case; details for the $N=5$ configuration are provided in the Supplemental 
Material~\cite{SM}.

\begin{figure*}
    \centering
    \includegraphics[width=1.0\textwidth]{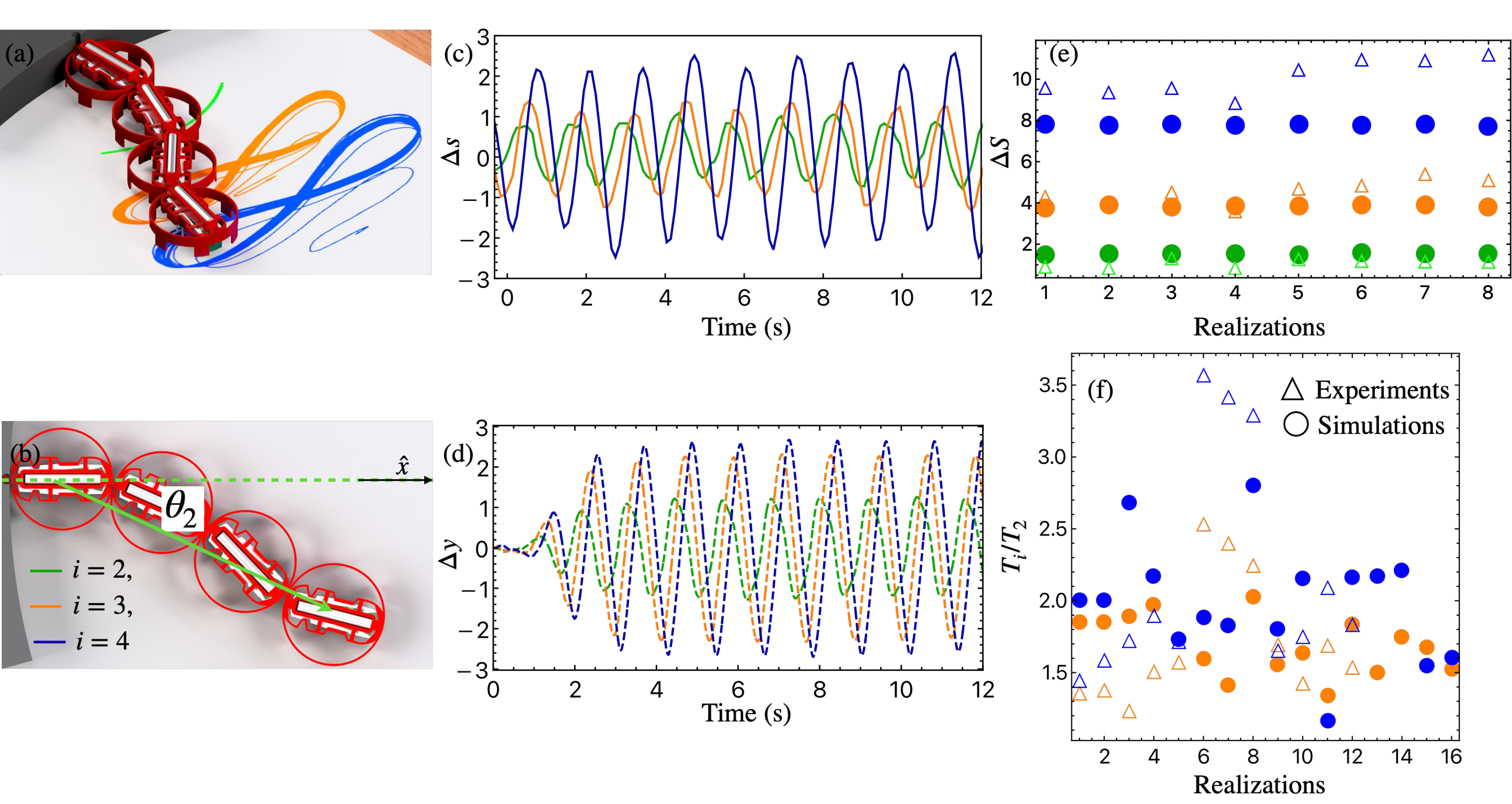}
    \caption{\textbf{Flagellar beating of a $N=4$ MSPP chain.} 
(a) Experimental photograph showing the trajectories of individual 
MSPPs during flagellar beating. Colors indicate particle index: green 
for $i=2$, orange for $i=3$, and blue for $i=4$ (tail); the head 
particle ($i=1$) is anchored and not shown. 
(b) Schematic of the $N=4$ anchored chain defining the oscillation 
angle $\theta_2$ between the position vector of the second particle 
and the chain axis $\hat{\mathbf{x}}$ (green dashed line). Red circles 
indicate the excluded-volume diameter $\sigma$ of each MSPP. 
(c) Arc displacement $\Delta s_i = (i-1)\sigma\theta_i(t)$ as a 
function of time from experiments. 
(d) Transverse displacement $\Delta y_i(t)$ as a function of time 
measured during numerical simulations. In both (c) and (d), amplitude increases 
with distance from the anchored head, consistent with flagellar 
beating~\cite{ma2014active,chelakkot2014flagellar}. 
(e) Oscillation amplitude $\Delta S_i = (i-1)\sigma A_i$ across 
multiple realizations. Filled circles represent simulations and 
triangles represent experiments. Amplitudes are consistent across 
realizations, indicating they are set by intrinsic system 
parameters. 
(f) Normalized oscillation period $T_i/T_2$ across multiple 
realizations. Periods vary between realizations, reflecting sensitivity 
to initial orientations and rotational noise at the moment of buckling.}
    \label{fig2}
\end{figure*}

We break the symmetry of the magnetic chain by anchoring the head particle 
to a 3D-printed wall that fixes its position ($\mathbf{r}_1=(0,0,0)$) and 
orientation ($\hat{\mathbf{p}} = -\hat{\mathbf{x}}$), forming a magnetically 
bound chain via dipole--dipole interactions (see Figure~\ref{fig2}). Due to the Hexbug's internal 
propulsion mechanism, the head exhibits minimal residual vibrations. All 
particles are activated simultaneously at the start of each experiment.

The motion of the chain is recorded using a standard cellphone camera. 
Post-processing was conducted using a custom Python code to extract the 
trajectories and orientations of individual particles (see Supplemental 
Material~\cite{SM}).

\subsection{The onset of buckling}
When particles are self-propelled along $\hat{\mathbf{p}} = -\hat{\mathbf{x}}$, 
head-to-tail magnetic dipoles generate attractive forces and torques that 
bind the particles together, while active forces introduce a longitudinal 
pressure $\Sigma_{\rm long} = F_0 N/\sigma^2$ that increases with chain 
length $\ell_{\rm chain} = N\sigma$, where $\sigma$ is the particle diameter 
and $F_0$ is the active force (see Fig.~\ref{fig1}).

This stress accumulation generates a symmetry breaking between the head 
and the tail of the chain. Whether buckling occurs depends on the 
competition between this accumulated stress and the magnetic bending 
stiffness $k_{\rm mag} = \mu_0 m^2/2\pi\sigma^2$. Following Euler buckling 
theory~\cite{nishiguchi2018flagellar,camalet1999self,chelakkot2014flagellar}, 
we define a critical force $F_c = k_{\rm mag}/\ell_{\rm chain}^2$ and a 
total active force $F_{\rm act} = F_0 N$. Their ratio defines the 
dimensionless buckling parameter
\begin{equation}
    \Pi_{\rm buck} = \frac{F_{\rm act}}{F_c} = 
    \frac{F_0\,N^3\sigma^2}{k_{\rm mag}},
    \label{eq:pibuck}
\end{equation}
which sets the balance between the active force $F_0$, the magnetic 
interaction strength $k_{\rm mag}$, and the chain length $N$. When 
$\Pi_{\rm buck}\ll 1$ the chain remains straight; when $\Pi_{\rm buck}\sim 1$ 
flagellar-like beating develops; and when $\Pi_{\rm buck}\gg 1$ fission is 
possible~\cite{kaiser2015active,guzman2016fission} (see phase diagram in 
Supplemental Material~\cite{SM}).

Substituting the experimental parameters ($\sigma = 0.05$~m, 
$m = 0.6$~Am$^2$, $F_0 =2\times 10^{-4}$~N) gives 
$k_{\rm mag} = \mu_0 m^2/2\pi\sigma^2 = 2.88\times10^{-5}$~N\,m$^2$ and
\begin{equation}
    \Pi_{\rm buck} = \frac{F_0\sigma^2}{k_{\rm mag}}\,N^3 
    = 1.736\times10^{-2}\,N^3.
\end{equation}
The resulting values for each chain length are summarized in 
Table~\ref{tab:pibuck}:

\begin{table}[h]
\centering
\begin{tabular}{ccc}
\hline
$N$ & $\Pi_{\rm buck}$ & Observed behavior \\
\hline
3 & 0.47 & chain remains straight \\
4 & 1.11& stable flagellar beating \\
5 & 2.17 & stable flagellar beating \\
6 & 3.75 & chain breaks (fission) \\
\hline
\end{tabular}
\caption{Buckling parameter $\Pi_{\rm buck}$ for each chain length $N$, 
computed from Eq.~\eqref{eq:pibuck} using the experimental parameters. 
The criterion $\Pi_{\rm buck}\sim 1$ accurately separates the three 
observed regimes.}
\label{tab:pibuck}
\end{table}

The table confirms that the criterion $\Pi_{\rm buck}\sim 1$ cleanly 
separates the three regimes. When the magnetic interaction is kept 
constant, a chain of $N=3$ cannot reach the buckling threshold and 
remains straight (see Supplemental Movie~SM1). For $N=6$, the active force overcomes magnetic 
cohesion and the chain breaks shortly after buckling (see Supplemental Movie~SM4). Stable flagellar 
beating is therefore observed only for $N=4$ and $N=5$, which bracket 
the critical value $\Pi_{\rm buck}\approx 1$ (see Supplemental Movie~SM2, SM3).

The dimensionless criterion $\Pi_{\rm buck}\sim 1$ is the discrete-chain 
analog of buckling parameters previously introduced for continuous active 
filaments~\cite{chelakkot2014flagellar,jayaraman2012autonomous,nishiguchi2018flagellar}, 
where the magnetic dipole-dipole interaction provides the effective 
bending rigidity $k_{\rm mag}$ directly from particle properties, 
enabling independent experimental control of both activity and stiffness.

\begin{figure*}[ht]
    \centering 
\includegraphics[width=1.0\textwidth]{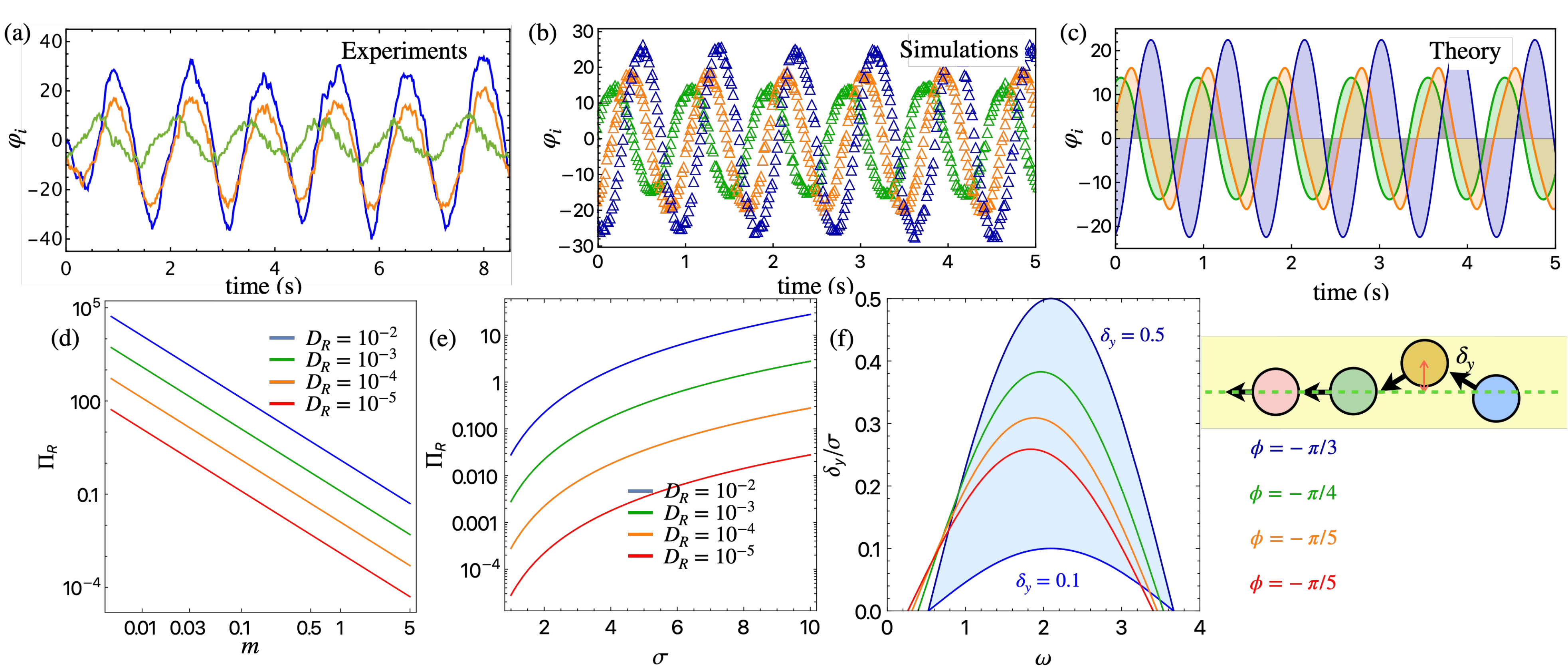}	
    \caption{\textbf{Orientation dynamics and onset of flagellar beating.}
(a) Time evolution of the orientation angle $\varphi_i$ for each 
particle during experiments. 
(b) Same quantity from numerical simulations (triangles). 
(c) Theoretical orientation time evolution from Eq.~\eqref{eq5} for 
$m=1$, $\sigma=1$, $\delta_y=0.45$, $\omega=0.72$, and $\phi=-\pi/4$ for each particle, with amplitude growing 
with distance from the anchored head. Colors in (a)--(c) indicate 
particle index: green for $i=2$, orange for $i=3$, and blue for $i=4$ 
(tail). 
(d) Rotational noise parameter $\Pi_R = 4\pi\gamma_R D_R\sigma^3/\mu_0 
m^2$ as a function of the magnetic dipole strength $m$ for fixed 
$\sigma$, shown for different rotational diffusion coefficients $D_R$. 
Larger $D_R$ (blue) leads to $\Pi_R \gg 1$, where noise dominates 
and stable flagellar beating is disrupted; smaller $D_R$ (red, as in 
our setup with $D_R = 10^{-5}$) gives $\Pi_R \ll 1$, allowing 
magnetic torques to maintain chain cohesion. 
(e) $\Pi_R$ as a function of particle diameter $\sigma$ for fixed $m$. 
Increasing $\sigma$ weakens the dipole-dipole interaction, raising 
$\Pi_R$ and reducing chain stability, consistent with the observation 
that larger particles require stronger dipole moments to sustain 
flagellar beating. 
(f) Stability diagram for the kinematic model: $\delta_y/\sigma$ as a 
function of angular frequency $\omega$ for different phase angles 
$\phi$, showing the range of $(\delta_y, \omega, \phi)$ values that 
sustain flagellar beating. The shaded region is bounded below by 
$\delta_{y,\rm min} = 0.1\sigma$ and above by $\delta_{y,\rm max} = 
0.5\sigma$. Dashed lines indicate values observed in experiments 
(gray) and simulations (red). Inset: schematic of the transverse 
displacement $\delta_y$ imposed on the tail particles to initiate 
beating.} 
\label{fig3}
\end{figure*}

\subsection{Flagellar beating} % o "Self-oscillations"

We performed numerical simulations modeling MSPPs as disk-shaped particles 
with point dipoles located at their centers. The position of the $i$-th 
particle is $\mathbf{r}_i(t) = (x_i(t), y_i(t), 0)$ and its orientation 
is $\hat{\mathbf{p}}_i(t) = (\cos\varphi_i(t), \sin\varphi_i(t), 0)$. 
The magnetic dipole moment is $\mathbf{m}_i = m\hat{\mathbf{p}}_i(t)$, 
where $m$ is its magnitude. Once buckling is triggered, the chain develops 
stable flagellar beating. The equations of motion are
\begin{eqnarray}
M\ddot{\mathbf{r}}_i &=& F_{0}\hat{\mathbf{p}}_i(t) - \gamma_{T}\dot{\mathbf{r}}_i 
- \nabla_{\mathbf{r}_i}\sum_{j\neq i}\left(U^{\text{WCA}}_{ij}+U^{\text{D}}_{ij}\right), 
\label{eq1}\\
\dot{\hat{\mathbf{p}}}_i &=& \frac{1}{\gamma_R}\left(\boldsymbol{\xi}_{i,R}(t)
-\mathbf{T}_i\right)\times\hat{\mathbf{p}}_i, \label{eq2}
\end{eqnarray}
where $\gamma_T$ and $\gamma_R$ are the translational and rotational 
friction coefficients, respectively, and $F_0$ is the active force. The 
term $\boldsymbol{\xi}_{i,R}(t)$ is a rotational Gaussian white noise of 
zero mean satisfying $\langle\xi_{j,R}(t_1)\rangle = 0$ and 
$\langle\xi_{j,R}(t_1)\xi_{j,R}(t_2)\rangle = 2\gamma_R^2 D_R\,\delta(t_1-t_2)$, 
where $D_R$ is the rotational diffusion coefficient. The magnetic torque 
acting on particle $i$ is $\mathbf{T}_i = \hat{\mathbf{p}}_i\times
\nabla_{\hat{\mathbf{p}}_i}U^{\text{D}}$.

We note that a self-alignment torque $\beta(\hat{\mathbf{p}}_i\times
\hat{\mathbf{v}}_i)$, which couples orientation to velocity, has been 
included in previous models of active magnetic 
particles~\cite{kaiser2015active,guzman2016fission,obreque2024dynamics} 
and shown to influence rotational dynamics in the absence of magnetic 
interactions. In the present system, however, magnetic torques dominate 
the orientational dynamics, and we verified that self-alignment plays no 
significant role in the flagellar beating observed here. Furthermore, 
since we study anchored chain configurations with fixed initial 
conditions, and since our analytical model (Eq.~\eqref{eq5}) reproduces 
the observed dynamics without this term, we omit it throughout.

Unlike colloidal systems typically modeled in the overdamped limit, 
Hexbug-based MSPPs operate at the macroscopic scale where inertial 
effects are non-negligible. The mass $M$ therefore appears explicitly in 
Eq.~\eqref{eq1}, placing this system in the regime of inertial magnetic 
active matter~\cite{obreque2024dynamics,musacchio2025fluidization, leoni2020surfing}, where translational dynamics retains inertia while orientational dynamics remains overdamped.

Each MSPP is subject to two interactions: an excluded-volume potential 
$U^{\text{WCA}}$ modeled by the Weeks--Chandler--Anderson (WCA) 
potential, and a magnetic dipole--dipole interaction $U^{\text{D}}$, 
given by
\begin{equation}
U^{\text{WCA}}_{ij} = \begin{cases}
4\varepsilon\!\left[\left(\dfrac{\sigma}{r_{ij}}\right)^{12}
-\left(\dfrac{\sigma}{r_{ij}}\right)^{6}\right] & r_{ij} \leq r_m,\\[6pt]
0 & \text{otherwise,}
\end{cases}
\label{eq:wca}
\end{equation}
\begin{equation}
U^{\text{D}}_{ij} = \frac{\mu_0 m^2}{4\pi r_{ij}^3}
\left[\hat{\mathbf{p}}_i\cdot\hat{\mathbf{p}}_j
-\frac{3(\hat{\mathbf{p}}_i\cdot\mathbf{r}_{ij})
(\hat{\mathbf{p}}_j\cdot\mathbf{r}_{ij})}{r_{ij}^2}\right],
\label{eq:dipole}
\end{equation}
where $\mathbf{r}_{ij} = \mathbf{r}_j - \mathbf{r}_i$, $r_{ij} = 
|\mathbf{r}_{ij}|$, $r_m = 2^{1/6}\sigma$ is the WCA cutoff, and 
$\varepsilon = 100\,k_BT$ is the excluded-volume energy scale.

For $N=4$ MSPPs (and $N=5$ in the Supplemental Material~\cite{SM}), with 
parameters $m=0.6$~Am$^2$ and $F_0=2\times10^{-4}$~N, we observe in 
both experiments and simulations that each particle in the chain undergoes 
flagellar beating (see Supplementary Movies~S2, S3, S5). The beating is 
characterized by the angle $\theta_i$, defined between the position 
vector of the $i$-th particle's center and the horizontal axis (see 
Fig.~\ref{fig2}(b)). Each particle follows a closed orbit whose amplitude 
increases with distance from the clamped head, as observed in sperm 
flagella and elastoactive 
simulations~\cite{ma2014active,guasto2020flagellar,zheng2023self}. This 
motion can be quantified by the arc length $\Delta s_i = (i-1)\sigma\theta_i$ 
(valid when $\theta_i$ is small) or equivalently by the transverse 
displacement $\Delta y_i$, where $i = 1, 2, \ldots, N$ and $i=1$ 
corresponds to the head particle.

We plot both $\Delta y$ and $\Delta s$ as a function of time in 
Fig.~\ref{fig2}(c),(d) for experiments and simulations, respectively. 
Stable periodic oscillations are observed, with amplitude increasing with 
distance from the clamped head. This behavior has been reported in 
previous experimental realizations in colloidal and inertial 
systems~\cite{nishiguchi2018flagellar,zheng2023self} and is consistently 
reproduced in our simulations. To quantify the oscillations, we fit a 
sinusoidal function $\Delta y_i(t) \approx \Delta s_i(t) = A_i\sin(2\pi 
t/T_i)$ to each particle's trajectory, extracting the oscillation 
amplitude $A_i$ and period $T_i$ across multiple experimental and 
numerical realizations.

To analyze and compare these results, we plot the arc swept 
$\Delta s_i = (i-1)\sigma A_i$ and the normalized period ratio $T_i/T_2$ 
for particles $i=3$ (orange) and $i=4$ (blue) in Fig.~\ref{fig2}(e),(f). 
Experimental data are shown as triangles and simulation results as 
circles, with all periods normalized by $T_2$, the period of the second 
particle in the chain.

Interestingly, the oscillation amplitudes remain consistent across 
realizations, indicating that they are set by intrinsic physical 
parameters such as the dipolar interaction strength $m$ and the active 
force $F_0$~\cite{chelakkot2014flagellar,guzman2016fission}. In contrast, 
the oscillation periods vary across realizations. On average, we measure 
$\langle T_3/T_2\rangle = 1.6$ and $\langle T_4/T_2\rangle = 1.9$, with 
the last particle generally exhibiting a longer period than its neighbors, 
a trend observed consistently in both simulations and experiments. This 
variability suggests that the oscillation frequency is not deterministic 
but is sensitive to rotational noise and the relative initial orientation 
of each particle at the moment of buckling.

We further investigate the role of magnetic interactions in shaping the 
flagellar beating.
\subsection{A simple theory for tangential misalignment}

During the experiments, we tracked the orientation angle $\varphi_i$ of 
each particle. Initially, all particles are aligned along the 
$-\hat{\mathbf{x}}$ direction, forming a straight magnetic chain. Upon 
activation, a small misalignment between neighboring particles, induced 
by the accumulation of longitudinal stress and rotational 
diffusion~\cite{chelakkot2014flagellar}, triggers buckling and the onset 
of flagellar beating. In the absence of rotational diffusion---as modeled 
in simulations---the system fails to break the initial symmetry, and even 
strongly active particles remain confined to a straight chain 
configuration~\cite{guzman2016fission}.

The orientation time evolution is shown in Fig.~\ref{fig3}(a),(b), where 
the amplitude of angular oscillation increases with the particle's 
distance from the anchored head. The angular displacements are limited by 
dipole--dipole magnetic torques, which tend to restore tangential 
alignment between neighboring dipoles. In Eq.~\eqref{eq2}, this torque 
counteracts rotational noise, acting as a restoring torque. The balance 
between the dipolar restoring torque and rotational noise defines a 
characteristic dimensionless misalignment parameter
\begin{equation}
   \Pi_R = 
    \frac{4\pi\gamma_R D_R\sigma^3}{\mu_0 m^2},
    \label{eq:delta}
\end{equation}
where $\Pi_R = 4\pi\gamma_R D_R\sigma^3/\mu_0 m^2$ measures the typical noise-induced angular misalignment between 
neighboring MSPPs. When $\Pi_R \ll 1$, magnetic torques dominate and the 
chain remains well-aligned; when $\Pi_R \gtrsim 1$, rotational noise 
overcomes magnetic cohesion and stable flagellar beating is disrupted.

In Fig.~\ref{fig3}(d), we plot $ \Pi_R$ as a function of $m$ for 
different $D_R$ values; red curves correspond to low $D_R \sim 10^{-5}$, 
while blue curves correspond to high $D_R \sim 10^{-2}$. For large $D_R$, 
$ \Pi_R > 1$ for most values of $m$, meaning rotational noise 
dominates and stable flagellar beating cannot be maintained. For small 
$D_R$, as in our setup ($D_R = 10^{-5}$), the noise-induced misalignment 
is small ($ \Pi_R \approx 0.006$ at $m = 0.6$~Am$^2$), allowing 
magnetic torques to sustain a stable chain configuration from which 
flagellar beating can develop once $\Pi_{\rm buck} \sim 1$.

In Fig.~\ref{fig3}(e), we show $ \Pi_R$ as a function of $\sigma$ 
for fixed $m$. As $\sigma$ increases, the dipole--dipole interaction 
weakens ($U^D \propto \sigma^{-3}$) while the noise-induced misalignment 
grows ($ \Pi_R \propto \sigma^{3}$), eventually exceeding the 
stability threshold. This confirms that larger particles require stronger 
dipole moments to sustain flagellar beating, consistent with our 
experimental observations.

To better understand the onset of buckling, we developed a theoretical 
model describing the time evolution of the orientation angle $\varphi_i$ 
due to magnetic torques,
\begin{equation}
  \dot{\varphi}_i = \frac{\mu_0 m^2}{4\pi\gamma_R} \sum_{j \neq i} 
  \hat{\mathbf{p}}_i \times 
  \left( \frac{3 \vec{\mathbf{r}}_{ij} (\hat{\mathbf{p}}_j \cdot 
  \vec{\mathbf{r}}_{ij})}{r_{ij}^5} 
  - \frac{\hat{\mathbf{p}}_j}{r_{ij}^3} \right).
  \label{eq5}
\end{equation}
We assume particle orientations $\hat{\mathbf{p}}_1 = (1, 0, 0)$ and, 
for $i=2,3,4$, $\hat{\mathbf{p}}_i = (\cos\varphi_i(t), 
\sin\varphi_i(t), 0)$, with positions $\vec{\mathbf{r}}_1 = (0, 0, 0)$, 
$\vec{\mathbf{r}}_2 = (\sigma, 0, 0)$, $\vec{\mathbf{r}}_3 = (2\sigma, 
\delta_y \cos(\omega t), 0)$, and $\vec{\mathbf{r}}_4 = (3\sigma, 
\delta_y \cos(\omega t + \phi), 0)$, where $\delta_y$ is the amplitude 
of the transverse oscillation imposed on particles 3 and 4, $\omega$ is 
an angular frequency, and $\phi$ a phase. Note that $\delta_y$ is the 
steady-state beating amplitude used as input to the kinematic model. When 
$\delta_y = 0$ the chain remains stable; for $\delta_y \neq 0$, flagellar 
beating emerges depending on $\phi$ and $\omega$. In Fig.~\ref{fig3}(c), 
filled curves show solutions to Eq.~\eqref{eq5} for $m=1$, $\sigma=1$, 
$\delta_y=0.45$, $\omega=0.72$, and $\phi=-\pi/4$, showing excellent 
agreement with simulations.

We find that a range of $(\delta_y, \phi)$ values leads to sustained 
flagellar beating for fixed $\sigma$ and $m$, as shown in 
Fig.~\ref{fig3}(f). The colored lines show $\delta_y/\sigma$ as a 
function of $\omega$ for different phase angles $\phi$, bounded below by 
the minimum amplitude required to sustain beating, 
$\delta_{y,\rm min}=0.1\sigma$, and above by the maximum, 
$\delta_{y,\rm max}=0.5\sigma$.

\begin{figure*}[ht!]
    \centering
    \includegraphics[width=0.8\textwidth]{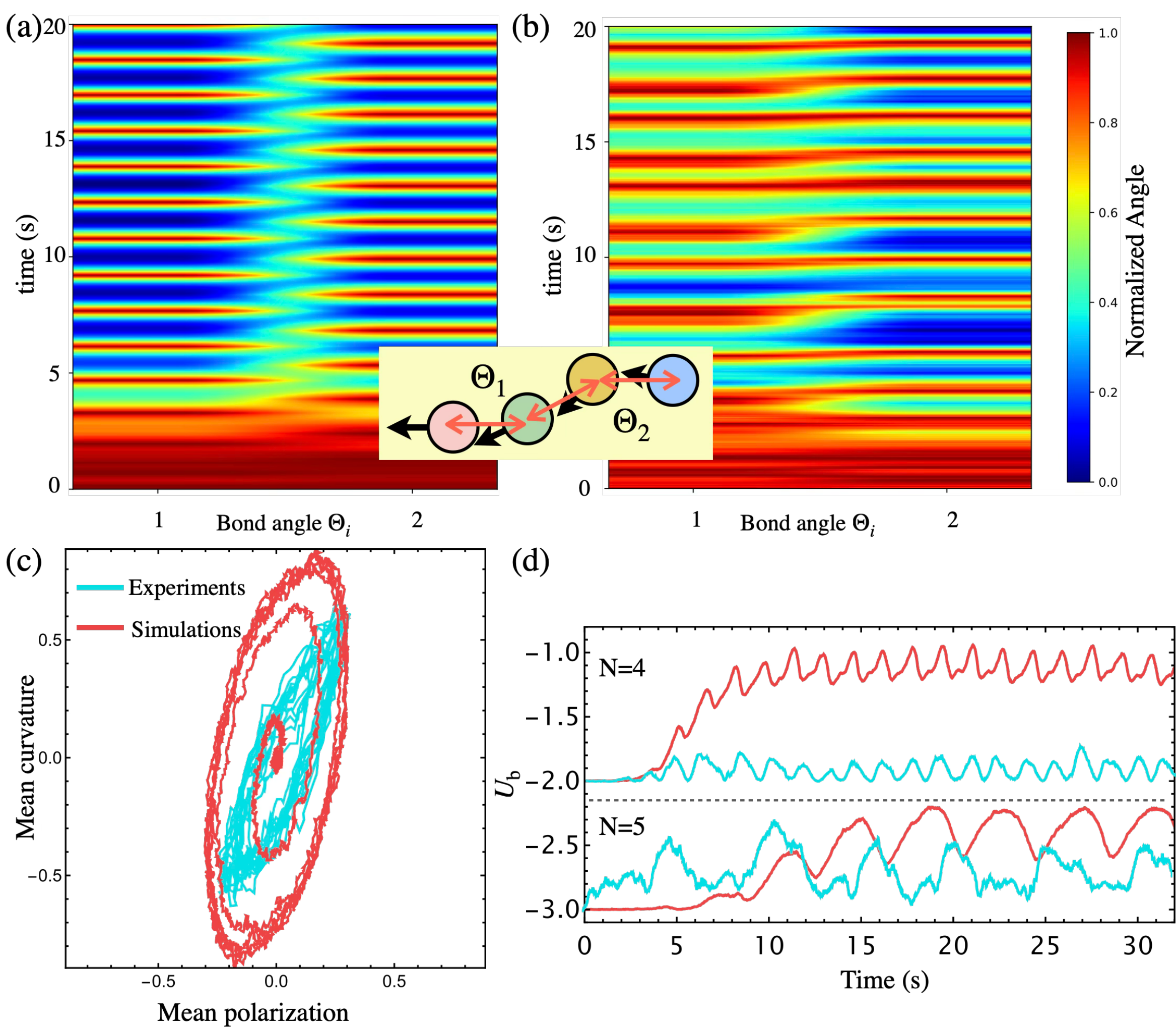}
    \caption{\textbf{Chain flexibility and limit cycle dynamics.}
(a),(b) Kymographs showing the temporal evolution of the bond angles 
$\Theta_1$ and $\Theta_2$ (defined in the schematic inset) for 
simulations (a) and experiments (b), respectively. Colors represent 
the bond angle normalized by the maximum deflection detected, from 
blue (minimum) to red (maximum). The periodic banding confirms 
sustained flagellar beating, with simulations showing more regular 
oscillations than experiments due to the absence of environmental 
noise.
(c) Phase space portrait of the mean curvature $\Phi(t)$ versus 
mean polarization $\Omega(t)$, showing the limit cycle structure 
characteristic of a supercritical Hopf 
bifurcation~\cite{zheng2023self,ling2018instability}. Experimental 
trajectories (cyan) exhibit larger variability than simulations (red), 
reflecting sensitivity to rotational noise and initial conditions.
(d) Time evolution of the three-body bending energy $U_b$ for 
flagella composed of $N=4$ (top) and $N=5$ (bottom) particles, from 
simulations (red) and experiments (cyan). The $N=5$ chain exhibits 
larger oscillation amplitudes and more negative mean $U_b$, consistent 
with its greater flexibility as predicted by the $k_{\rm mag}/(N\sigma)^2$ 
scaling~\cite{chelakkot2014flagellar,li2010bending}.}
    \label{fig4}
\end{figure*}
\subsection{How soft is the magnetic flagellum?}

The onset of these oscillations corresponds to a \textit{supercritical 
Hopf bifurcation}~\cite{zheng2023self,ling2018instability,de2017spontaneous}, 
leading to a limit cycle in terms of the system's mean polarization 
$\Omega(t) = \frac{1}{N}\sum_{i=1}^N \varphi_i(t)$ and mean bond angle 
$\Phi(t) = \frac{1}{N-1}\sum_{i=1}^{N-1}\Theta_i(t)$, where 
$\Theta_i = \angle(\mathbf{b}_i, \mathbf{b}_{i+1})$ is the angle between 
consecutive bond vectors $\mathbf{b}_i = \mathbf{r}_{i+1} - \mathbf{r}_i$, 
as shown in Fig.~\ref{fig4}(a) (red for experiments, gray for simulations). 
This limit cycle structure is analogous to that observed in elastoactive 
structures~\cite{zheng2023self} and beating biological 
flagella~\cite{ma2014active,guasto2020flagellar}, where mean curvature 
and polarization serve as natural order parameters for the oscillatory 
state.

Next, we investigate the flexibility of the magnetic flagellum. The 
magnetic bending stiffness per particle pair, $\kappa_{\rm mag} = 
\mu_0m^2/2\pi\sigma^2$, sets the resistance to relative angular 
deflection between neighboring MSPPs~\cite{kiani2015elastic,li2010bending}. 
The effective rigidity of the chain as a whole, however, is governed by 
the Euler critical force $F_c = \kappa_{\rm mag}/(N\sigma)^2$, which 
decreases as $N^{-2}$ with chain 
length~\cite{camalet1999self,chelakkot2014flagellar}. Consequently, 
shorter chains are stiffer and harder to buckle, while longer chains are 
more flexible and buckle under weaker active forcing, consistent with our 
experimental observations (see Supplementary Movies~SM1--SM4).

To quantify this flexibility, we analyze the bond angles $\Theta_i$, 
defined in the schematic of Fig.~\ref{fig4}, and compute kymographs 
showing their time evolution for simulations and experiments in 
Fig.~\ref{fig4}(b),(c), respectively. Colors represent the bond angle 
normalized by the maximum deflection detected in each realization.

We model the chain bending using a three-body potential based on the 
unit bond vectors $\mathbf{b}_i = (\mathbf{r}_i - 
\mathbf{r}_{i-1})/|\mathbf{r}_i - \mathbf{r}_{i-1}|$, yielding the 
total bending 
energy~\cite{li2010bending,chelakkot2014flagellar,laskar2015brownian}
\begin{equation}
    U_b = \frac{k_{\rm mag}}{2} \sum_{i=2}^{N-1} 
    \left( \mathbf{b}_{i+1} - \mathbf{b}_i \right)^2,
    \label{bendingpot}
\end{equation}
where $k_{\rm mag} = \mu_0m^2/2\pi\sigma^2$ is the magnetic bending 
stiffness, which here plays the role of the bending rigidity $\kappa$ 
in the discrete worm-like chain model~\cite{li2010bending}. Unlike 
passive polymer chains where $\kappa$ is fixed by molecular architecture, 
here $k_{\rm mag}$ is tunable via the dipole moment $m$, offering direct 
experimental control over chain flexibility~\cite{kiani2015elastic}. 
The bending energy over time is shown in Fig.~\ref{fig4}(d),(e), where 
oscillation amplitudes grow with $N$ in agreement with the predicted 
$k_{\rm mag}/(N\sigma)^2$ scaling~\cite{tejedor2024progressive}.

Our work provides a physical realization of beating by 
propulsion~\cite{camalet1999self,jayaraman2012autonomous} in a model 
system where both bending rigidity and activity arise from controllable, 
experimentally tunable parameters. By combining magnetic dipolar 
interactions with self-propelled motion in centimeter-scale robots 
(Hexbugs)~\cite{obreque2024dynamics,musacchio2025fluidization}, we 
construct magnetic self-propelled particles (MSPPs) that self-assemble 
into flexible chains and undergo spontaneous flagellar beating when one 
end is clamped.

Unlike microscale systems where flagellar motion typically results from 
internal motor activity or elastic deformations in slender 
filaments~\cite{de2017spontaneous,chelakkot2014flagellar,laskar2015brownian}, 
here the oscillations are driven purely by the interplay between active 
forces, magnetic cohesion, and orientational fluctuations. Elasticity 
emerges effectively from the magnetic dipole-dipole interactions, giving 
rise to a bending rigidity $k_{\rm mag}=\mu_0m^2/2\pi\sigma^2$ that can 
be modulated via dipole strength~\cite{kiani2015elastic,li2010bending}, 
while the effective chain flexibility grows with particle number as 
$F_c = k_{\rm mag}/(N\sigma)^2$, consistent with Euler buckling 
theory~\cite{camalet1999self,chelakkot2014flagellar,nishiguchi2018flagellar}. 
The onset of beating is initiated by a critical tangential misalignment 
$\delta$ that triggers buckling, and rotational diffusion plays an 
essential role in symmetry breaking — without it, as shown in 
simulations, the chain remains in a metastable straight configuration 
regardless of propulsion 
strength~\cite{guzman2016fission,chelakkot2014flagellar}.

By systematically analyzing both experiments and simulations, we 
demonstrate that flagellar beating is robust and reproducible, with 
amplitude and frequency depending nonlinearly on dipolar strength and 
chain length~\cite{nishiguchi2018flagellar,zheng2023self}. The 
macroscopic nature of our model offers unique advantages: trajectories 
and orientations can be tracked with high spatiotemporal resolution using 
simple imaging tools~\cite{obreque2024dynamics}; propulsion strength, 
dipolar coupling, and chain architecture can be modified in a controlled 
manner; and long timescales allow direct observation of transient and 
steady-state dynamics, including transitions between linear and nonlinear 
regimes~\cite{zheng2023self,tejedor2024progressive}.

The observed flagellar beating is reminiscent of locomotion in 
spermatozoa and cilia~\cite{gilpin2020multiscale,cammann2025form,guasto2020flagellar}, 
suggesting that the underlying principles — force-induced buckling 
coupled to orientational dynamics — may be generic across 
scales~\cite{ling2018instability,ma2014active}. While our current system 
focuses on a single flagellum, future studies could explore collective 
effects in systems of multiple interacting chains, where flagellar 
synchronization via magnetic or elastic coupling may give rise to 
behaviors analogous to those observed in biological arrays of cilia and 
flagella~\cite{wan2024mechanisms,moreau2024minimal,xia2024biomimetic}. 
Our system further provides a platform for testing reduced models of 
active filaments, including symmetry-breaking bifurcations, limit cycles, 
and mode selection in nonlinear 
oscillators~\cite{zheng2023self,ling2018instability,de2017spontaneous} 
— offering insights directly transferable to synthetic 
microswimmers~\cite{vutukuri2017rational,gonzalez2018active} and 
biological systems governed by active forces and mechanical constraints.

\section*{Acknowledgements}
We thank Professor Eric Clement for helpful discussions.
N. Q, D.H and F.G.-L. have received support from Fondecyt Regular No.\ 1250913.
\vspace{0.1cm}

\section*{Author contributions}
 All authors contributed equally to the analysis and writing of the manuscript; F.G.-L, E.L. and E.B. designed the research, F.G.-L. performed the simulations. N.Q, D.H and E.B performed the experiments. N.Q. and D.H wrote the scripts to analyze the experimental data.

\section*{Additional information}
There are no conflicts to declare.

\bibliography{Referencias}

\appendix

\end{document}